\documentclass[aps,prl,twocolumn,showpacs,preprintnumbers,superscriptaddress,amsmath,amssymb]{revtex4-1}
\usepackage{latexsym}
\usepackage{graphicx}
\usepackage{graphics}
\usepackage[T1]{fontenc}
\usepackage{multirow}
\usepackage{xcolor, soul}
\sethlcolor{pink}
\usepackage{textcomp, gensymb}
\usepackage{lineno}

\begin{document}


\author{Jianpei Geng} \email{jianpei.geng@pi3.uni-stuttgart.de} \thanks{These authors contributed equally.}
\author{Tetyana Shalomayeva} \email{tetyana.shalomayeva@pi3.uni-stuttgart.de} \thanks{These authors contributed equally.}
\author{Mariia Gryzlova}
\author{Amlan Mukherjee}
\author{Santo Santonocito}
\author{Dzhavid Dzhavadzade}
\author{Durga Dasari}
\affiliation{3rd Institute of Physics, ZAQuant, University of Stuttgart, 70569 Stuttgart, Germany}
\author{Hiromitsu Kato}
\affiliation{National Institute of Advanced Industrial Science and Technology (AIST), Tsukuba, Ibaraki 305-8568, Japan}
\author{Rainer St{\"o}hr} 
\author{Andrej Denisenko} 
\affiliation{3rd Institute of Physics, ZAQuant, University of Stuttgart, 70569 Stuttgart, Germany}
\author{Norikazu Mizuochi}
\affiliation {Institute for Chemical Research, Kyoto University, Gokasho, Uji, Kyoto, 611-0011, Japan}
\affiliation{Center for Spintronics Research Network, Kyoto University, Uji, Kyoto, 611-0011, Japan}
\affiliation{International Center for Quantum-field Measurement Systems for Studies of the Universe and Particles (QUP), KEK, Tsukuba, Ibaraki 305-0801, Japan}
\author{J{\"o}rg Wrachtrup}
\affiliation{3rd Institute of Physics, ZAQuant, University of Stuttgart, 70569 Stuttgart, Germany}

\title{Dopant-assisted stabilization of negatively charged single nitrogen-vacancy centers in phosphorus-doped diamond at low temperatures}

\begin{abstract}
Charge state instabilities have been a bottleneck for the implementation of solid-state spin systems and pose a major challenge to the development of spin-based quantum technologies. Here we investigate the stabilization of negatively charged nitrogen-vacancy (NV$^-$) centers in phosphorus-doped diamond at liquid helium temperatures. Photoionization of phosphorous donors in conjunction with charge diffusion at the nanoscale enhances NV$^0$ to NV$^-$ conversion and stabilizes the NV$^-$ charge state without the need for an additional repump laser.
The phosphorus-assisted stabilization is explored and confirmed both with experiments and our theoretical model.
Stable photoluminescence-excitation spectra are obtained for NV$^-$ centers created during the growth. 
The fluorescence is continuously recorded under resonant excitation to real-time monitor the charge state and the ionization and recombination rates are extracted from time traces.
We find a linear laser power dependence of the recombination rate as opposed to the conventional quadratic dependence, which is attributed to the photo-ionization of phosphorus atoms.

\end{abstract}

\maketitle

\section{Introduction}
\begin{figure}[ht]
\includegraphics[width=1\columnwidth]{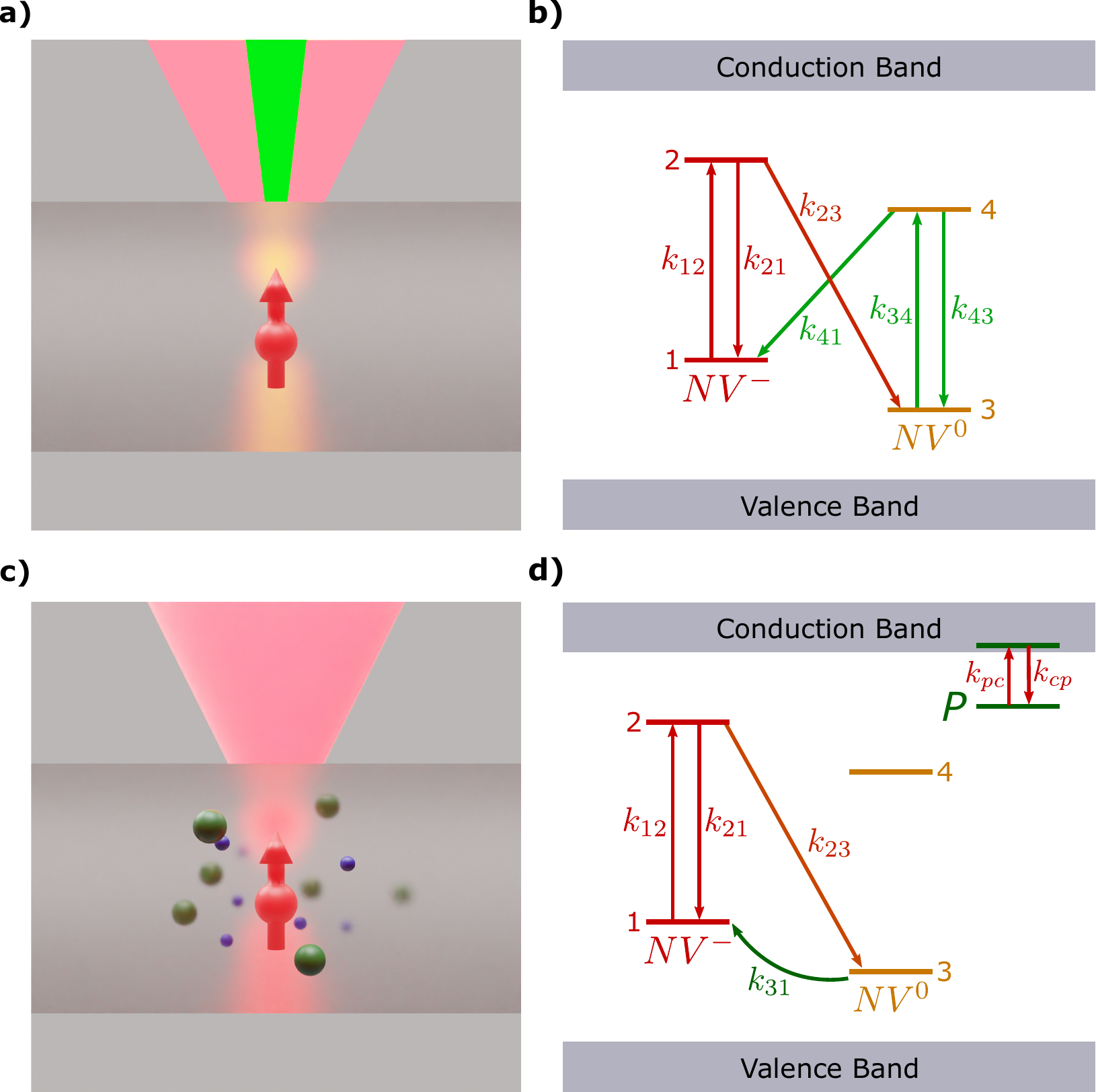}
\caption{\textbf{Charge state conversion of NV center in diamond.}
(a) Schematic representation of the conventional method to keep the NV center in NV$^-$ with repump laser. The red sphere with an arrow represents the NV center in diamond. The red and green cones represent the illumination of the red resonant and additional repump lasers respectively.
(b) Illustration of the NV photo- and charge dynamics under the illumination of the resonant and repump lasers. The red solid lines with arrows indicate the population transfer of the NV state due to the resonant laser, which ultimately ionizes the NV center into NV$^0$. The green solid lines with arrows indicate the recombination pathway of population transfer provided by the repump laser.
(c) Schematic representation of NV$^-$ charge state stabilization assisted by photo-ionization of phosphorus impurities. The red sphere with an arrow represents the NV center in diamond. The red cone represents the illumination of the red laser resonant to the NV$^-$ excitation. The green spheres show phosphorus impurities. The blue spheres represent electrons created by the photo-ionization of the phosphorus impurities.
(d) Illustration of the photo- and charge dynamics of the NV center in phosphorus-doped diamond under the illumination of the resonant laser. The energy levels of both the NV center and the phosphorus impurity are presented. The red solid lines with arrows indicate the photo-ionization of the NV center and the phosphorus impurity due to the resonant laser excitation. The green solid line indicates the recombination pathway provided by capturing the electrons created by the photo-ionization of the phosphorus impurities.
}
\label{fig1} 
\end{figure}

Spin defects in wide-bandgap semiconductors \cite{DOHERTY20131, christle2015isolated, Nathan2021hBN} are promising candidates for quantum information processing due to their excellent magneto-optical properties. The negatively charged nitrogen-vacancy (NV$^-$) center in diamond \cite{DOHERTY20131} has become a leading contender among these defects benefiting from efficient optical initialization and readout of the spin state \cite{Neumann2010NucRead, Hanson2011NVRead}, long spin coherence time \cite{balasubramanian2009ultralong}, and high-fidelity quantum control \cite{Du2015NVGate}. 
Substantial improvement has been achieved towards using these defects for spin-based quantum computation\cite{Neumann2010NVRegister, Taminiau2022NVLogicalQubit}, simulation \cite{Du2019PT, Taminiau2021DTC}, nanoscale quantum sensing and imaging \cite{Lukin2008NVSensing, Joerg2008NVImaging, Du2015SingleProtein, Qichao2021MagneticDomain}, and quantum networks \cite{Lukin2010SpinPhotonEntanglement, Hanson2013SeparateNVEntanglement, Hanson2021Network, Hanson2022QubitTeleportation}. 
 As only  the negatively charged NV$^-$ state  is useful for the aforementioned applications, a key challenge is to continuously monitor its charge state and stabilize it. 
 The stabilization of the NV$^-$ charge state becomes even more important for shallow NV centers \cite{PhysRevLett.122.076101, D2TC01258H} wherein sensitivity and spatial resolution get compromised in quantum sensing and imaging experiments \cite{Lukin2008SensitivityResolution, Awschalom2015ProtonMRI} due to charge instability.
 Usually, a repump mechanism is invoked using an additional laser to make sure that the defect
remains in its correct charge state for further processing as shown in Fig. \ref{fig1}a and b.
This additional check and the involved repump mechanisms have been a bottleneck and slow down the performance of these systems for cutting-edge industrial applications, and also increase the overhead in miniaturization trends.
The use of a repump laser also causes spurious effects such as  spectral diffusion \cite{Hanson2010NVOpticalTransition} deteriorating  the spin-photon interfaces\cite{Faraon2012CoupleNVCavity} formed with these defects.
Conventionally, this repump laser is used in between the operating pulses to bring the defect back to the correct charge state i.e, a conversion from the neutrally charged NV$^0$ to the negatively charged NV$^-$ state \cite{aslam2013photo, PhysRevLett.110.167402, wirtitsch2023exploiting}.
Here we overcome this limitation by introducing dopants in the diamond lattice acting as electron donors thereby intrinsically assisting this charge state conversion without the need for external repump fields.

Among the various methods \cite{D2TC01258H, Garrido2011ChemicalChargeControl, Garrido2012ElectricalChargeControl, Mizuochi2014ElectricalNVCharge, Jan2014Doping, PhysRevB.93.081203, herbschleb2019ultra, Mizuochi2021ShallowNVinPDiamond} developed to control and stabilize the charge state of NV centers in diamond, a promising one is doping of electron donor impurities such as phosphorus \cite{Jan2014Doping, PhysRevB.93.081203, herbschleb2019ultra, Mizuochi2021ShallowNVinPDiamond}. Previous studies of charge state dynamics of NV centers in phosphorus-doped diamond either used ensemble samples or room temperature experiments, where a relative NV$^-$/NV$^0$ population has been analyzed under the  illumination of the green or orange laser. While these studies clearly indicate the charge state stabilization by photo-ionized electrons from the dopants, a clear distinction between this stabilization from that caused by a repump laser is not evident. This is due to the fact that the excitation lasers used in their experiment can also cause the NV$^0$ - NV$^-$ transition alongside those caused by the photoionized electrons.
Hence a quantitative description through its experimental verification of how the phosphorus impurities stabilize the NV$^-$ charge state at the level of a single center is essential to understand the physical mechanism leading to such charge stabilization. Exploring these effects at low temperatures of $4K$ opens up the venue for its application in a large number of quantum protocols. For this, we here monitor the charge-state dynamics under resonant excitation for NV centers in phosphorus-doped diamond at cryogenic temperatures and show how the photo-ionization of the phosphorous donors affects the recombination/transition rates among the charge states and explore the differences from the repump mechanism using a repump-laser. 

In this letter, we investigate the charge state stability of NV$^-$ centers in phosphorus-doped diamond under resonant excitation at liquid helium temperatures and reveal the key role of photo-ionization of phosphorus impurities in the charge-state stabilization of the defect. Consequently, stable photoluminescence-excitation (PLE) spectra of NV$^-$ centers created during the growth process are obtained without any repump laser. This enables us to realize real-time monitoring of the NV charge state by recording fluorescence under 636 nm resonant excitation. Ionization and recombination rates of the NV centers are extracted from time traces of the fluorescence at various excitation powers. Further, the linear power dependence of the recombination rate was observed, distinguished from the well-known quadratic dependence \cite{aslam2013photo} for NV centers in the intrinsic diamond when a repump laser is applied. This linear power dependence could be attributed to the dependence of the photo-ionization induced carrier density on the laser power. 
Finally, a theoretical model is developed to gain insight into the charge state stabilization of NV centers assisted by photo-ionization of the phosphorus impurities, which is helpful in the optimization of phosphorus concentration to realize stable and narrow PLE spectra of NV$^-$ centers created during the growth process.

\section{Results}

\begin{figure}[ht]
\includegraphics[width=1\columnwidth]{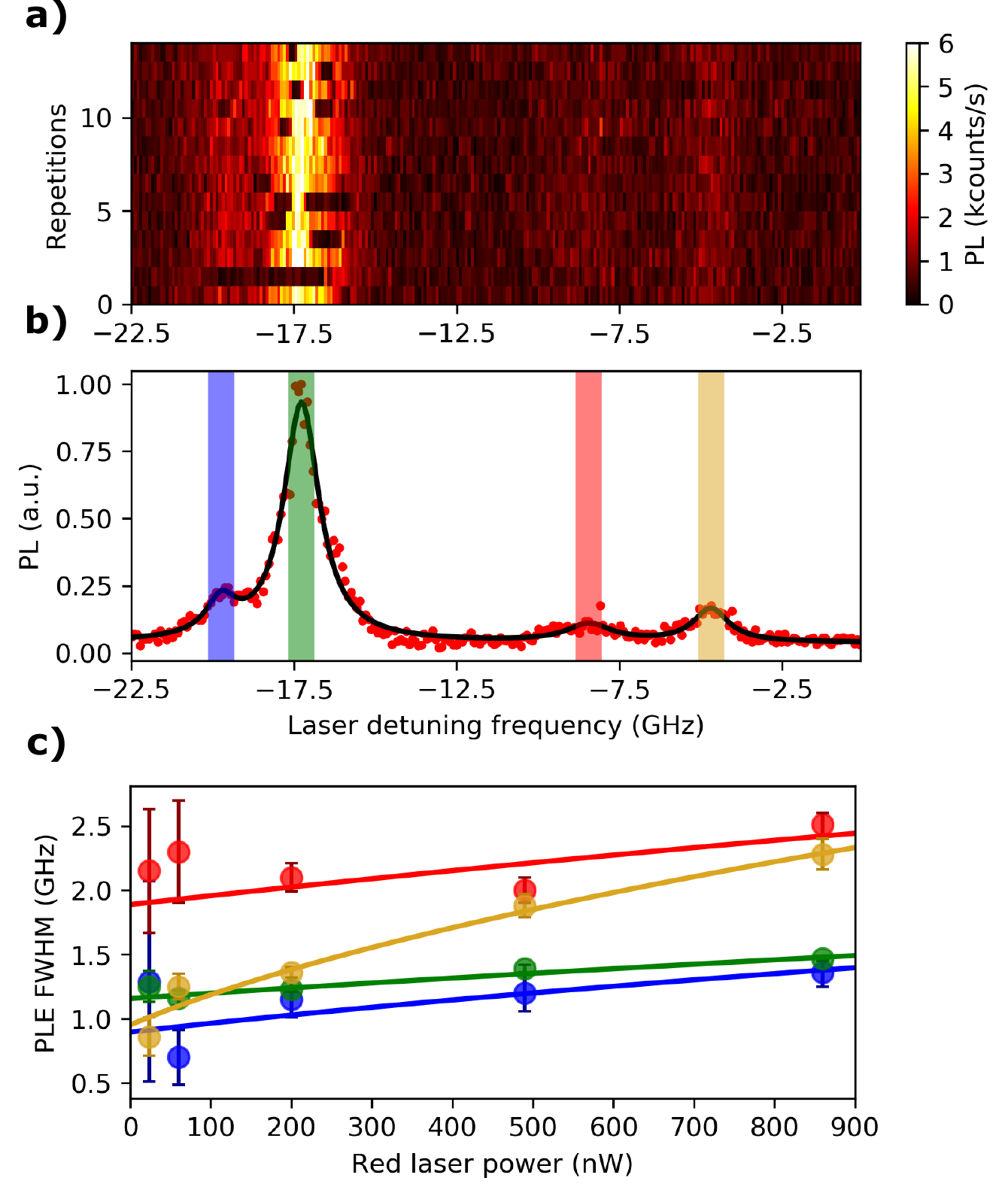}
\caption{\textbf{Stable PLE spectrum of an NV center created during growth in the phosphorus-doped diamond.}
(a) PL intensity as a function of laser detuning (detuned from 636.60 nm) in several individual repetitions without repumping. 
Sharp changes in the PL intensity indicate NV ionization or recombination events. The measurements show a stable but broad PLE spectrum of NV$^-$ over multiple repetitions.
(b) PLE spectrum averaged over multiple repetitions in (a). The circles are experimental data and the solid line is fit with a multi-Lorentz function. Blue, green, red, and yellow marked regions correspond to 4 different observed peaks.
(c) Resonant laser power dependence of PLE full width at half maximum (FWHM). Colors represent the FWHM values for each individual peak in (b) respectively. The fit is shown by solid lines.
}
\label{fig2} 
\end{figure}
\subsection*{Experiment}
The measurements were performed on NV centers in a phosphorus-doped diamond with a home-built cryogenic confocal setup. The phosphorus-doped layer was epitaxially grown onto Ib-type (111)-oriented diamond substrate by CVD with phosphorus concentration of $5\times10^{16}$ atoms $\text{cm}^{-3}$. NV centers created during growth could be found in the phosphorus-doped layer. Additional NV centers were created by implantation of $^{15}$N$^+$ ions in the phosphorus-doped layer with an energy of 9.8 keV and a dose of $1.3\times10^{10}$ atoms $\text{cm}^{-2}$. Arrays of nanopillars \cite{momenzadeh2015nanoengineered} with apex sizes of 400, 450, and 500 nm were fabricated to enhance the photon collection efficiency of NV centers created by ion implantation. The NV centers created during growth and those created by ion implantation could be distinguished from their position, depth, and from the hyperfine spectrum seen in the ODMR signal. We performed measurements on both types of NV centers for comparison and to gain deep insight into the role of phosphorus impurities in the NV spin and charge-state properties. The average spin coherence (Hahn Echo) time at liquid helium temperatures was measured to be (i) $T_2=1.94$ ms for NV centers created during growth and (ii) $T_2=11.53$ $\mu$s for those created by ion implantation. The long spin coherence time of NV centers created during growth is attributed to the suppression of paramagnetic vacancy complexes due to their charging by phosphorus atoms\cite{herbschleb2019ultra}. 
Further details on the experimental setup, sample preparation, and spin coherence time measurements are given in the Supplementary Information. 

Figure \ref{fig2} shows the PLE measurements of an NV center created during the growth in a phosphorus-doped diamond at liquid helium temperature. The measurements were performed by recording the fluorescence while sweeping the laser frequency across the NV$^-$ resonance ($637$nm). A peak in the PLE spectrum is observed when the NV$^-$ center gets resonantly excited. However, since the resonant excitation could photo-ionize NV$^-$ into NV$^0$ and not excite NV$^0$ back into NV$^-$ \cite{PRL109_097404PLER}, the NV center would essentially stay in NV$^0$ charge state leading to the disappearance of the PLE peak. A recombination pathway from NV$^0$ to NV$^-$ requires an additional electron which becomes available by an additional laser i.e., optical repumping either with a 488 nm \cite{PRL109_097404PLER}, 532 nm \cite{Hanson2010NVOpticalTransition}, or 575 nm laser \cite{PhysRevLett.110.167402}. 
On the other hand, as shown in Fig. \ref{fig2} we observe a stable PLE spectrum of the NV$^-$ center even in the absence of the repump laser. The disappearance of fluorescence in an individual PLE scan indicates the ionization of NV$^-$ into NV$^0$, and the fluorescence recovery indicates the recombination of NV$^0$ to NV$^-$. Due to this continuous ionization and recombination processes, we observe a stable PLE spectrum of all measured NV centers created during growth in the phosphorus-doped layer of diamond without the need for a repump laser (see Supporting Information). This clearly indicates that a recombination pathway has been provided by the phosphorus impurities. 

Phosphorus is a shallow electron donor in diamond with an energy level located at 0.57 eV below the conduction band edge \cite{Sque2004ShallowDonors}, and thus could be photo-ionized by the 636 nm laser used in the PLE measurements. Electrons created by photo-ionization of the phosphorus impurities could be captured by NV$^0$, providing the recombination pathway. Figure \ref{fig2}b shows the PLE spectrum of an NV$^-$ center averaged over multiple repetitions, which is fitted by a multi-Lorentz function. The curve fitting gives a linewidth of 1 GHz for the observed PLE spectrum and is much broader than the Fourier transform limited linewidth of NV centers in intrinsic diamond \cite{PhysRevLett.97.083002}. The broadening is attributed to the environmental charge fluctuation induced by photo-ionized electrons of the phosphorus impurities. We measured PLE spectra at various excitation powers, and the power dependence of the PLE linewidths is shown in Fig. \ref{fig2}c. 

\begin{figure}[ht]
\includegraphics[width=1\columnwidth]{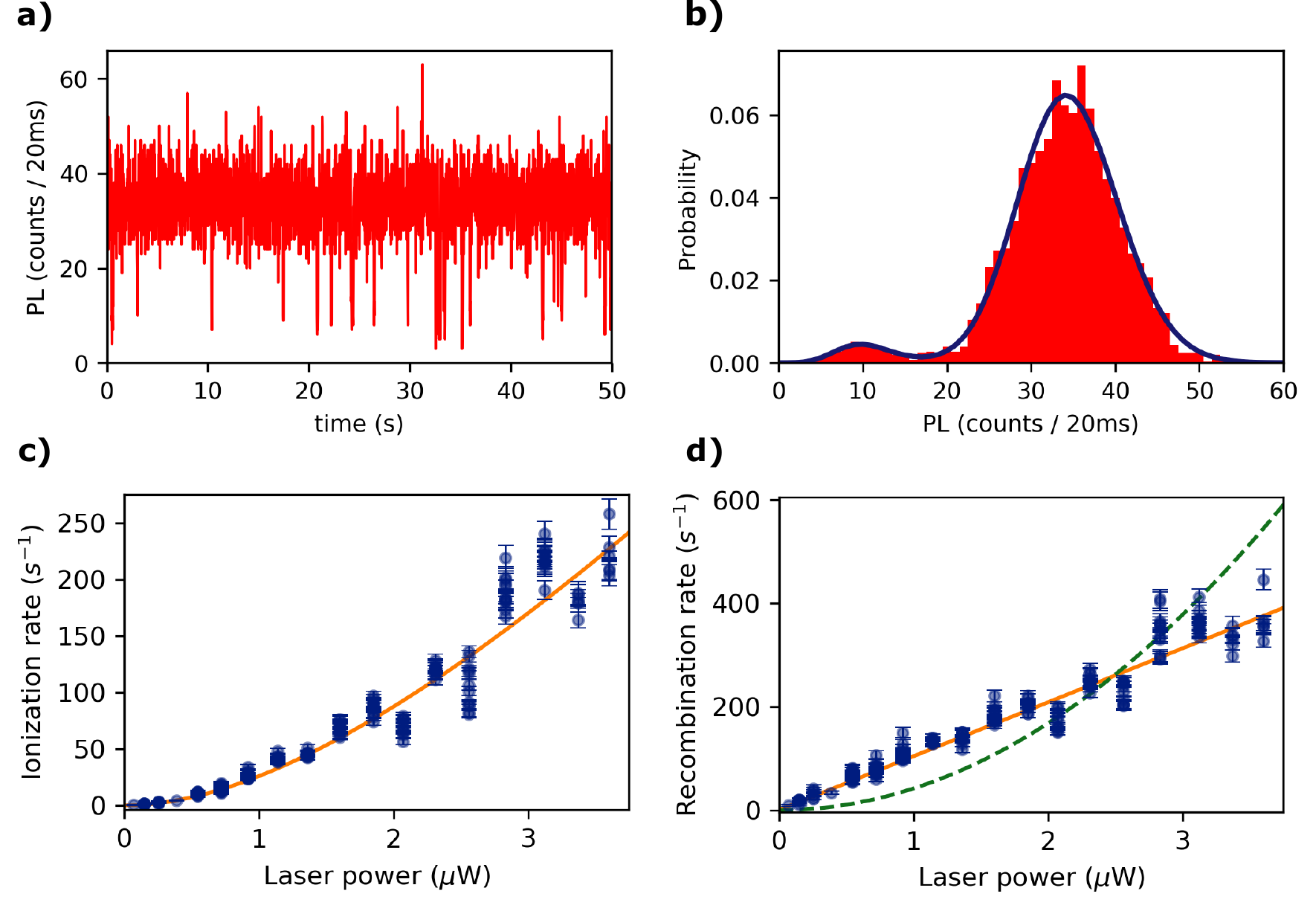}
\caption{\textbf{Charge state population and laser power dependence of rates in charge dynamics of an NV center created during growth in the phosphorus-doped diamond.}
(a) Time trace of the fluorescence under continuous illumination of 148 nW, 636 nm resonant laser. Blinking events arise from dynamic NV charge ionization and recombination. 
(b) Histogram of (a), showing a larger population of NV$^-$ than NV$^0$. The solid line is a fit.
(c) Laser power dependence of ionization rate. The circles represent experimental data and the solid line is a fit with a saturation-modified quadratic function. The saturation power is fixed as \( 4.5\ \mu \)W in the fitting.
(d) Laser power dependence of recombination rate. The circles are experimental data and the solid line is a linear fit according to the model. For comparison, the dashed line is a quadratic fit according to the laser power dependence for NV centers in the non-doped diamond. 
}
\label{fig3} 
\end{figure}

The charge dynamics of the NV centers that are created during growth is investigated under the resonant excitation of a 636 nm laser. We would like to note that at this frequency only the NV$^-$ charge state is excited but not the NV$^0$ i.e., fluorescence from the NV center can be detected only when it is in the NV$^-$ charge state. Therefore, the charge state could be real-time monitored by recording the fluorescence. If the excitation is spin selective, driving the spin transitions in the NV$^-$ ground state by an appropriate microwave field would suppress any spin polarization \cite{Hanson2011NVRead} and maintain the NV$^-$ fluorescence. In our experiment, a microwave drive was not necessary, because all the spin levels could be simultaneously excited by a single excitation as the PLE lines are quite broad. Figure \ref{fig3}a shows a typical time trace of the fluorescence where the resonant excitation power is chosen to be 148 nW. The high count level indicates the NV center in NV$^-$ charge state and the low level is the background counts with the NV center in NV$^0$. As we can see, the NV center stays for a much shorter time in NV$^0$ than in NV$^-$, corresponding to a larger recombination rate than the ionization rate. The intensity histogram of the time trace is presented in Fig. \ref{fig3}b, showing a relatively larger population of NV$^-$ than the NV$^0$ charge state. The histogram is fitted with the function derived by Shields \textit{et al.} \cite{Lukin2015SpinChargeConversion}, in which the photon count distribution is described by count rates for NV$^-$ and NV$^0$ charge states together with the ionization and recombination rates determining the charge-state switching. A set of distributions of the photon count within various counting times from the same time trace is fitted to improve the fitting precision of the rates (see Supplementary Information). The fitting result is shown as the solid line in Fig. \ref{fig3}b, which gives an ionization rate of $k_{\text{ion}}=0.5(1)$ s$^{-1}$ and a recombination rate of $k_\text{rec}=11(3)$ s$^{-1}$, and thus an NV$^-$ population of $P_-=k_\text{rec}/(k_{\text{ion}}+k_\text{rec})=96(2)\%$. The time traces of fluorescence are recorded at various excitation powers, and corresponding photon count distributions are fitted to extract the rates. Figure \ref{fig3}c and d show the excitation power dependence of the ionization and recombination rate, respectively. The ionization rate could be well fitted with function $k_{\text{ion}}=aP_\text{L}^2/(1+P_\text{L}/P_\text{s})$, where $P_\text{L}$ is the excitation power, $P_\text{s}$ is the saturation power, and $a$ is a fitting coefficient. The good agreement of the numerical fit to the obtained experimental data indicates a two-photon process of the photo-ionization of NV$^-$, which is consistent with previous research \cite{aslam2013photo, PhysRevLett.110.167402}. However, the recombination rate shows a linear excitation power dependence which is in contrast to the previous observations where a similar two-photon excitation pathway has been known. This linear dependence could be attributed to the additional electrons provided by the photoionized phosphorus atoms, and in the following, we present a simple model to explain this power dependence.

\subsection{Model}

As mentioned before, the resonant excitation is absorbed only by NV$^-$ and hence alone cannot contribute to the recombination pathway from  NV$^0$ to  NV$^-$. Therefore, the recombination pathway could only be provided by the electrons from the donor impurities. The laser power dependence of the recombination rate indicates the interplay of the laser intensity and the concentration of phosphorus impurities. Photoconductivity studies of photoionized phosphorus impurities in diamond \cite{Gheeraert1999LTPhotoConduct, Koizumi2003PhotoHall} have clearly shown dependence on the excitation laser and impurity concentration. From these studies and our experimental observations, we develop below a simple model to extract the recombination rate and its dependence on the above-mentioned parameters.  As illustrated in Fig. \ref{fig1}c, we consider phosphorus impurities within the focal volume of the NV center excitation. Photo-ionized electrons created by the photo-ionization of the impurities could diffuse spatially and get captured by the NV center to convert it from NV$^0$ to NV$^-$. The spatial power distribution of the laser (beam profile) can be described as
\begin{equation}
    I(\vec r)=P_\text{L}f(\vec r),
    \label{Eq1}
\end{equation}
where $P_\text{L}$ is the laser power, $f(\vec r)$ is the normalized intensity distribution satisfying $\int f(\vec r)dS=1$, and $I(\vec r)$ is the laser intensity at position $\vec r$. The photo-ionization of the phosphorus impurity is modeled with the transition of the electron from the phosphorus level to the conduction band, as is shown in Fig. \ref{fig1}d. The transition rate $k_\text{pc}$ is proportional to the laser intensity, and is thus a function of $\vec r$,
\begin{equation}
    k_\text{pc}(\vec r)=\alpha I(\vec r),
    \label{Eq2}
\end{equation}
where $\alpha$ is determined by the photo-ionization cross-section of phosphorous atoms in a diamond lattice, and the absorption cross-section that can vary depending on the position of the defects in the sample. The rate of the electron transition from the conduction band back to the phosphorus level is denoted as $k_\text{cp}$. We can now approximate the electronic density (population) in the conduction band due to the photo-ionization of phosphorus impurities at position $\vec r$ as
\begin{equation}
    P_\text{e}(\vec r)=\frac{k_\text{pc}(\vec r)}{k_\text{pc}(\vec r)+k_\text{cp}}.
    \label{Eq3}
\end{equation}
The electron density $n_\text{e}$ at the position of the NV center due to diffusion of the electrons created by photo-ionization of the phosphorus impurities can be written as
\begin{equation}
    n_\text{e}=\int \eta(\vec r)P_\text{e}(\vec r) n_\text{P}dV,
    \label{Eq4}
\end{equation}
where $n_\text{P}$ is the phosphorus concentration, and $\eta (\vec r)$ is the probability that an electron in the conduction band at position $\vec r$ diffuses into a unit volume at the position of the NV center. We assume a homogeneous phosphorus concentration $n_\text{P}$ for simplicity. $\eta (\vec r)$ is dependent on the electron diffusion constant and lifetime due to capture by impurities. The capture of electrons by the NV$^0$ center gives the NV recombination rate
\begin{equation}
    k_\text{rec}=\kappa n_e,
    \label{Eq5}
\end{equation}
where $\kappa$ is the electron capture coefficient of NV$^0$. With Eq. \ref{Eq1}-\ref{Eq5}, $k_\text{rec}$ could be derived as
\begin{equation}
    k_\text{rec}=\kappa\alpha\int\frac{\eta(\vec r)f(\vec r)}{\alpha P_\text{L}f(\vec r)+k_\text{cp}}dVn_\text{P}P_\text{L}.
    \label{Eq6}
\end{equation}
The linear laser power dependence of the recombination rate would be immediately obtained if the laser power is small, i.e. $\alpha P_\text{L}f(\vec r)\ll k_\text{cp}$, which gives
\begin{equation}
    k_\text{rec}=\kappa\frac{\alpha}{k_\text{cp}}\int\eta(\vec r)f(\vec r)dVn_\text{P}P_\text{L},
    \label{Eq7}
\end{equation}
consistent with the experimental results.

\begin{figure}[ht]
\includegraphics[width=1\columnwidth]{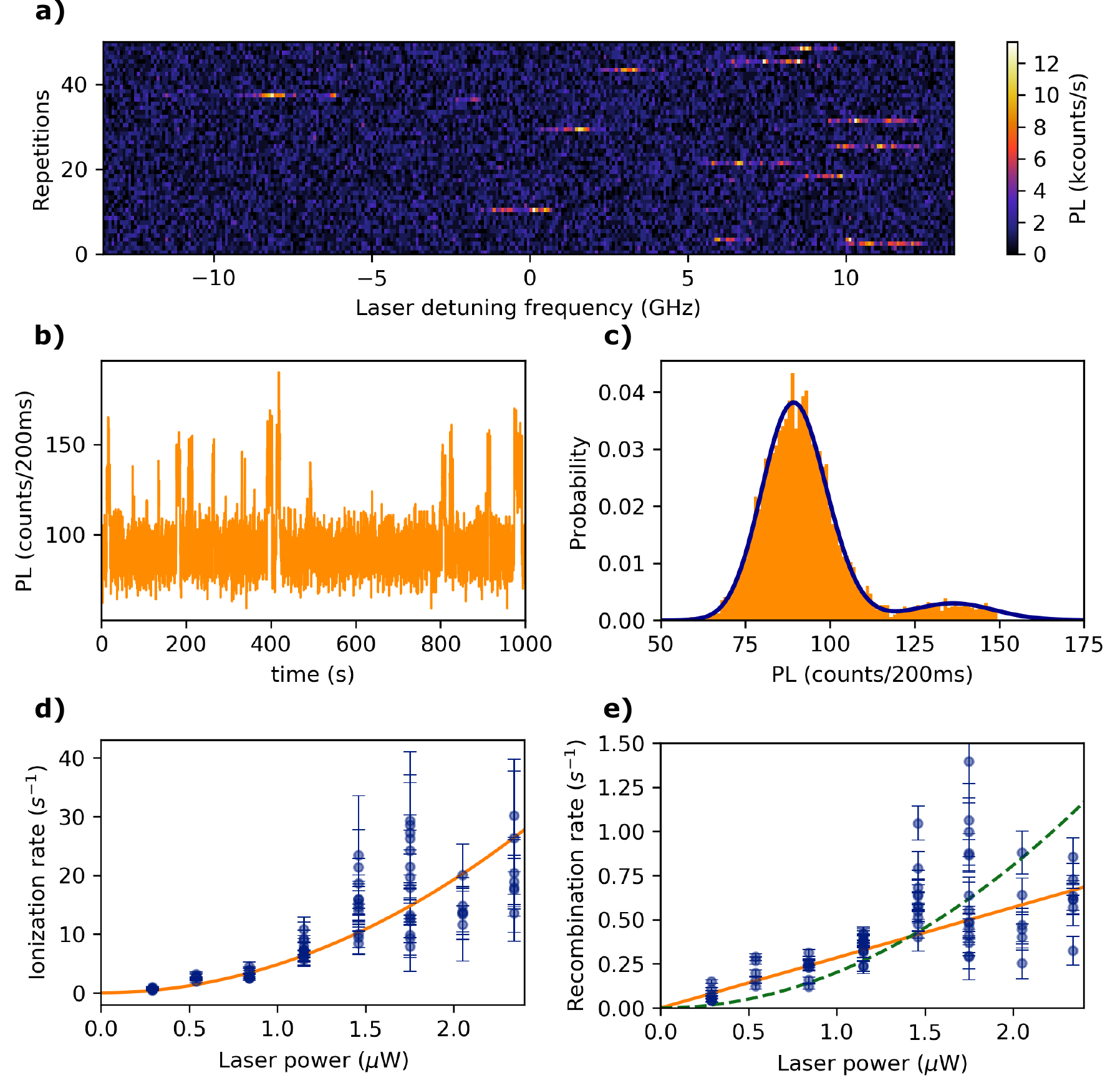}

\caption{\textbf{ PLE spectrum, charge state population, and laser power dependence of the charge dynamic rates of an NV center created by ion implantation in the phosphorus-doped diamond.}
(a) PL intensity as a function of laser detuning (detuned from 636.6 nm) in several individual repetitions without repumping. 
Unstable resonant excitation is observed only at certain individual repetitions, indicating the NV center in NV$^0$ charge state most of the time.
(b) Time trace of the fluorescence of the NV center under continuous illumination with 290 nW, 594 nm laser, showing NV$^-$ charge state with higher counts and NV$^0$ with lower counts. 
(c) Histogram of (b), showing a larger population of NV$^0$ than NV$^-$. The solid line is a fit.
(d)Laser power dependence of ionization rate. The circles represent experimental data and the solid line is a fit with a saturation-modified quadratic function. Each circle corresponds to an individual measurement.
(e)Laser power dependence of recombination rate. The circles are experimental data and the solid line is a linear fit according to the model. For comparison, the dashed line is a quadratic fit according to the laser power dependence for NV centers in the non-doped diamond. }
\label{fig4} 
\end{figure}

The model is further confirmed with measurements on NV centers created by ion implantation in the phosphorus-doped diamond. The implantation could create a large number of impurities such as vacancy-complex. The impurities play the role of either electron traps or electron acceptors which capture electrons from the valance band and create holes under the laser illumination. In each case, the probability $\eta(\vec r)$ decreases due to the trap or combination of the electrons by the impurities or holes. According to Eq. \ref{Eq7}, this causes a decrease in the recombination rate but doesn't change the linear laser power dependence. We performed PLE measurements on multiple NV centers created by ion implantation without repump laser. For many of them, the PLE spectra weren't detectable. For others, PLE peaks are observable only in certain repetitions as shown in Fig. \ref{fig4}a. In most repetitions of the PLE scan, the NV centers are in a dark state. These measurements indicate a much smaller recombination rate of NV centers created by ion implantation than those created during growth. Real-time monitoring of the charge state of these NV centers by recording fluorescence under resonant excitation becomes challenging due to the too small recombination rates. That's why we real-time monitor the charge state with 594 nm excitation. A typical time trace of the fluorescence and its histogram is shown in Fig. \ref{fig4}b and c, where a small but non-zero NV$^-$ population is observed. The extracted ionization and recombination rate as a function of laser power is shown in Fig. \ref{fig4}d and e, respectively. The ionization rate shows a quadratic power dependence, and the recombination rate could be fitted with a linear function of the laser power. These results coincide with the expectation from the model.

\section{Discussion}

While the photo-ionized electrons have a positive impact on the recombination rate they on the other hand could have a negative effect on the optical linewidth due to increased electric noise. Counterintuitively we have found that the line width is insensitive to the applied power of the resonant laser (Fig. \ref{fig2}c), and remained unchanged for increasing laser power up to a $\mu$W. This counterintuitive effect can be understood only upon properly evaluating the electric field noise taking into consideration the screening effects caused by the photo-ionized free charge carriers. 
Using the Thomas-Fermi model for free electron gas at low temperatures we find the electron density dependence of the screening length and further the modified electric field (see Supplementary Information for more details).
We find that the electric field shows a linear increase with increasing electron density $n_\text{e}$ when $n_\text{e}$ is small, and an exponential decay when $n_\text{e}$ is large.
At a certain value of $n_\text{e}$ the electric field reaches a maximum and remains almost unchanged for a considerable change of the electron density $n_\text{e}$ near this value.
From our experimental measurements, we conclude that we are in such a field-insensitive regime, from which we obtain the mean typical electron density to be $\sim 10^{10} - \sim 10^{14}$ cm$^{-3}$. This is in line with the actual phosphorous concentration of $10^{16}$ cm$^{-3}$, and assuming that not all the phosphorous dopants are ionized as explained above, an almost unchanged PLE linewidth dependence on the laser power can be understood.

\section{Conclusion}
In conclusion, we report PLE and charge-state measurements on single NV centers in phosphorus-doped diamond at $4K$. Stable PLE spectra without a repump laser, and a near unity probability to find the ingrown NV center in its correct charge state ($NV^-$) under resonant excitation are observed. A linear excitation power dependence of the recombination rate is observed on NV centers created during growth clearly distinguishing the dopant-assisted recombination pathway from the laser-assisted process. For defect centers created by ion implantation, while no stable PLE spectrum could be observed, a recombination pathway from NV$^0$ to NV$^-$ (with low probability) is observed due to the presence of phosphorous impurities. Here once again a linear laser power dependence of the recombination rate is detected indicating a dopant-assisted recombination process.  A simple model is developed to briefly describe various dependencies of the recombination rate on the laser and the dopant concentration. Further optimization of phosphorus concentration for stable and narrow PLE spectrum could be advantageous for quantum network applications as it removes the necessity of time-consuming repump which might enhance entanglement rates substantially \cite{Hanson2021Network, Hanson2022QubitTeleportation}.

\section{Methods}
\subsection{Experimental Setup}
The measurements were performed with a home-built cryogenic confocal microscope. The phosphorus-doped diamond with a phosphorus concentration of $5\times10^{16}$ cm$^{-3}$ was mounted inside a Janis ST-500 cold finger Helium flow cryostat at 4 K. The chamber was evacuated to $10^{-8}$ mbar. The lasers for excitation and phonon-sideband fluorescence from the NV centers went through the same air objective (Nikon LU Plan Fluor 100$\times$, NA0.9). Each laser passed through an acousto–optic modulator (AOM) twice for pulse generation before going into the objective. The fluorescence was separated from the excitation lasers with a dichroic mirror and a 660-800 nm bandpass filter, and then detected by Perkin-Elmer avalanche photodiodes (APDs). More details about the experimental setup could be found in the Supplementary Information.
 
\section{Data Availability}
The data that support the findings of this study are available from the corresponding authors upon reasonable request.

\section{Acknowledgements}
We acknowledge financial support by European Union’s Horizon 2020 research and innovation program AMADEUS under grant No. 101080136, Federal Ministry of Education and Research (BMBF) project MiLiQuant, Quamapolis, and Cluster4Future-QSens, as well as SPINNING under grant No.13N16219. NM would like to acknowledge the support by JSPS KAKENHI (21H04653), MEXT Q-LEAP (JPMXS0118067395), the Collaborative Research Program of Institute for Chemical Research, Kyoto University (2022-72), and the Spintronics Research Network of Japan.

\section{Author Contributions}
J.G. and T.S. contributed equally to this work. J.W., R.S., and J.G. designed the experiment. J.G., T.S., M.G., A.M., and D.D. (Dzhavadzade) conducted the experiment. J.G. and T.S. analyzed the data with support from M.G. J.G., T.S., and D.D. (Dasari) developed the model. J.G. made the numerical simulation. N.M. and H.K. prepared the phosphorus-doped diamond sample. A.D. and S.S. made the ion implantation. R.S. created the nanopillars. J.G., T.S., and D.D. (Dasari) wrote the manuscript. All authors discussed about the data, commented on the manuscript, and contributed to the paper.

\section{Competing Interests}
The authors declare no competing interests.

\bibliography{references.bib}



\end{document}